\documentclass[twocolumn]{revtex4}

\usepackage{bm}
\usepackage{amsthm,amsmath,amssymb}
\usepackage[dvips]{graphicx}

\theoremstyle{plain}
\newtheorem{theorem}{Theorem}

\newtheorem{proposition}[theorem]{Proposition}

\newcommand{\R}{{\mathbb R}}

\newcommand{\supp}{\mathop{\textup{supp}}}
\newcommand{\co}{\colon}
\newcommand{\qedbox}{\hfill$\Box$}

\begin{document}

\title{Algorithmic design of self-assembling structures
\footnotetext{Proc.\ Natl.\ Acad.\ Sci.\ USA, 2009, vol.~106, no.~24,
pp.~9570--9575.}}

\author{Henry Cohn}
\affiliation{Microsoft Research New England, One Memorial Drive,
Cambridge, MA 02142} \email{cohn@microsoft.com}

\author{Abhinav Kumar}
\affiliation{Department of Mathematics, Massachusetts Institute of
Technology, Cambridge, MA 02139} \email{abhinav@math.mit.edu}


\begin{abstract}
We study inverse statistical mechanics: how can one design a potential
function so as to produce a specified ground state?  In this paper, we
show that unexpectedly simple potential functions suffice for certain
symmetrical configurations, and we apply techniques from coding and
information theory to provide mathematical proof that the ground state
has been achieved. These potential functions are required to be
decreasing and convex, which rules out the use of potential wells.
Furthermore, we give an algorithm for constructing a potential function
with a desired ground state.
\end{abstract}

\maketitle

\section{Introduction}

How can one engineer conditions under which a desired structure will
spontaneously self-assemble from simpler components?  This inverse
problem arises naturally in many fields, such as chemistry, materials
science, biotechnology, or nanotechnology (see for example
ref.~\cite{To} and the references cited therein). A full solution
remains distant, but in this paper we develop connections with coding
and information theory, and we apply these connections to give a
detailed mathematical analysis of several fundamental cases.

Our work is inspired by a series of papers by Rechtsman, Stillinger,
and Torquato, in which they design potential functions that can produce
a honeycomb \cite{RST1}, square \cite{RST2}, cubic \cite{RST3}, or
diamond \cite{RST4} lattice.  In this paper, we analyze finite
analogues of these structures, and we show similar results for much
simpler classes of potential functions.

For an initial example, suppose $20$ identical point particles are
confined to the surface of a unit sphere (in the spirit of the Thomson
problem of how classical electrons arrange themselves on a spherical
shell).  We wish them to form a regular dodecahedron with $12$
pentagonal facets.

Suppose the only flexibility we have in designing the system is that we
can specify an isotropic pair potential $V$ between the points.  In
other words, the potential energy $E_V(\mathcal{C})$ of a configuration
(i.e., set of points) $\mathcal{C}$ is
\begin{equation}\label{eq:Edef}
E_V(\mathcal{C}) = \frac{1}{2}\sum_{x,y \in \mathcal{C},\, x \ne y}
V \big(|x-y|\big).
\end{equation}
In static equilibrium, the point configuration will assume a form that
at least locally minimizes $E_V(\mathcal{C})$.  Can we arrange for the
energy-minimizing configuration to be a dodecahedron?  Furthermore, can
we arrange for it to have a large basin of attraction under natural
processes such as gradient descent?  If so, then we can truly say that
the dodecahedron automatically self-assembles out of randomly arranged
points when the proper potential function is imposed.

\begin{figure}
\begin{center}
\includegraphics{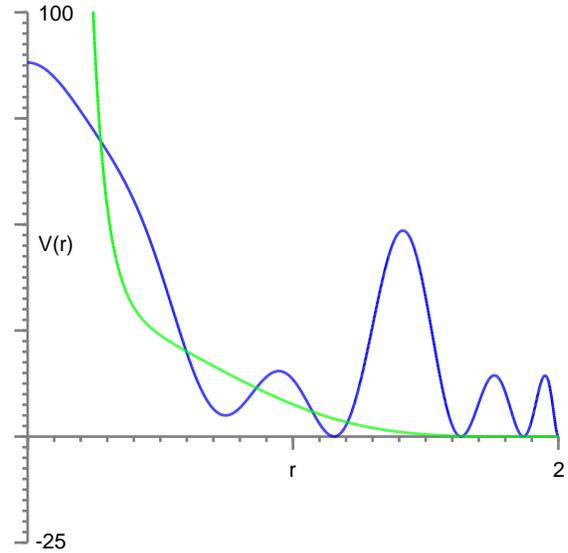}
\end{center}
\caption{Two potential functions under which the regular dodecahedron
minimizes energy: the blue one uses potential wells, while the green
one is convex and decreasing.}
\label{fig:potwellsconvex}
\end{figure}

If we could choose $V$ arbitrarily, then it would certainly be possible
to make the dodecahedron the global minimum for energy by using
potential wells, as in the blue graph in Fig.~\ref{fig:potwellsconvex}.
By contrast, this cannot be done with familiar potential functions,
such as inverse power laws, because the dodecahedron's pentagonal
facets are highly unstable and prone to collapse into a triangulation.

Unfortunately, the potential function shown in the blue graph in
Fig.~\ref{fig:potwellsconvex} is quite elaborate.  Actually
implementing precisely specified potential wells in a physical system
would be an enormous challenge.  Instead, one might ask for a simpler
potential function, for example one that is decreasing and convex
(corresponding to a repulsive, decaying force).

\begin{figure}
\begin{center}
\includegraphics{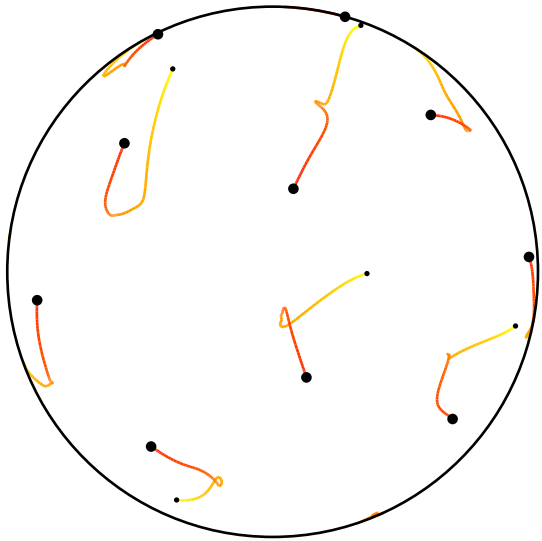}
\end{center}
\caption{The paths of points converging to the regular dodecahedron
under the green potential function from Fig.~\ref{fig:potwellsconvex}.
Only the front half of the sphere is shown.}
\label{fig:dodec-top}
\end{figure}

In fact, $V$ can be chosen to be both decreasing and convex.  The green
graph in Fig.~\ref{fig:potwellsconvex} shows such a potential function,
which is described and analyzed in Theorem~\ref{theorem:dodecahedron}.
We prove that the regular dodecahedron is the unique ground state for
this system.  We have been unable to prove anything about the basin of
attraction, but computer simulations indicate that it is large (we
performed $1200$ independent trials using random starting
configurations, and all but $6$ converged to the dodecahedron). For
example, Fig.~\ref{fig:dodec-top} shows the paths of the particles in a
typical case, with the passage of time indicated by the transition from
yellow to red.

Our approach to this problem makes extensive use of linear programming.
This enables us to give a probabilistic algorithm for inverse
statistical mechanics. Using it, we construct simple potential
functions with counterintuitive ground states.  These states are
analogues of those studied in refs.~\cite{RST1, RST2, RST3, RST4}, but
we use much simpler potential functions.  Finally, we make use of the
linear programming bounds from coding theory to give rigorous
mathematical proofs for some of our assertions. These bounds allow us
to prove that the desired configurations are the true ground states of
our potential functions. By contrast, previous results in this area
were purely experimental and could not be rigorously analyzed.

\subsection{Assumptions and Model}

To arrive at a tractable problem, we make four fundamental assumptions.
First, we will deal with only finitely many particles confined to a
bounded region of space.  This is not an important restriction in
itself, because periodic boundary conditions could create an
effectively infinite number of particles.

Second, we will use classical physics, rather than quantum mechanics.
Our ideas are not intrinsically classical, but computational necessity
forces our hand.  Quantum systems are difficult to simulate classically
(otherwise the field of quantum computing would not exist), and there
is little point in attempting to design systems computationally when we
cannot even simulate them.  Fortunately, classical approximations are
often of real-world as well as theoretical value.  For example, they
are excellent models for soft matter systems such as polymers and
colloids \cite{L,RSS}.

Third, we restrict our attention to a limited class of potential
functions, namely isotropic pair potentials.  These potentials depend
only on the pairwise distances between the particles, with no
directionality and no three-particle interactions; they are the
simplest potential functions worthy of analysis.  For example, the
classical electric potential is of this sort.  We expect that our
methods will prove useful in more complex cases, but isotropic pair
potentials have received the most attention in the literature and
already present many challenges.

Finally, we assume all the particles are identical.  This assumption
plays no algorithmic role and is made purely for the sake of
convenience.  The prettiest structures are often the most symmetrical,
and using identical particles facilitates such symmetry.

We must still specify the ambient space for the particles.  Three
choices are particularly natural: we could study finite clusters of
particles in Euclidean space, configurations in a flat torus (i.e., a
region in space with periodic boundary conditions, so the number of
particles is effectively infinite), or configurations on a sphere.  Our
algorithms apply to all three cases, but in this paper we will focus on
spherical configurations. They are in many ways the most symmetrical
case, and they are commonly analyzed, for example in Thomson's problem
of arranging classical electrons on a sphere.

Thus, we will use the following model. Suppose we have $N$ identical
point particles confined to the surface of the unit sphere $S^{n-1} =
\{x \in \R^n: |x|=1 \}$ in $n$-dimensional Euclidean space $\R^n$. (We
choose to work in units in which the radius is $1$, but of course any
other radius can be achieved by a simple rescaling.) We use a potential
function $V \co (0,2] \to \R$, for which we define the energy of a
configuration $\mathcal{C} \subset S^{n-1}$ as in Eq.~\eqref{eq:Edef}.
We define $V$ only on $(0,2]$ because no other distances occur between
distinct points on the unit sphere.

Note that we have formulated the problem in an arbitrary number $n$ of
dimensions.  It might seem that $n=3$ would be the most relevant for
the real world, but $n=4$ is also a contender, because the surface of
the unit sphere in $\R^4$ is itself a three-dimensional manifold
(merely embedded in four dimensions). We can think of $S^3$ as an
idealized model of a curved three-dimensional space.  This curvature is
important for the problem of ``geometrical frustration'' \cite{SM}:
many beautiful local configurations of particles do not extend to
global configurations, but once the ambient space is given a small
amount of curvature they piece together cleanly. As the curvature tends
to zero (equivalently, as the radius of the sphere or the number of
particles tends to infinity), we recover the Euclidean behavior.
Although this may sound like an abstract trick, it sometimes provides a
strikingly appropriate model for a real-world phenomenon; see, for
example, figure~2.6 in ref.~\cite{SM}, which compares the radial
distribution function obtained by X-ray diffraction on amorphous iron
to that from a regular polytope in $S^3$ and finds an excellent match
between the peaks.

The case of the ordinary sphere $S^2$ in $\R^3$ is also more closely
connected to actual applications that it might at first appear.  One
scenario is a Pickering emulsion, in which colloidal particles adsorb
onto small droplets in the emulsion.  The particles are essentially
confined to the surface of the sphere and can interact with each other,
for example, via a screened Coulomb potential or by more elaborate
potentials.  This approach has in fact been used in practice to
fabricate colloidosomes \cite{DHNMBW}. See also the review article
\cite{BG}, in particular section~1.2, and the references cited therein
for more examples of physics on curved, two-dimensional surfaces, such
as amphiphilic membranes or viral capsids.

\subsection{Questions and Problems}

From the static perspective, we wish to understand what the ground
state is (i.e., which configuration minimizes energy) and what other
local minima exist.  From the dynamic perspective, we wish to
understand the movement of particles and the basins of attraction of
the local minima for energy.  There are several fundamental questions:

\begin{enumerate}
\item Given a configuration $\mathcal{C}$, can one choose an
    isotropic pair potential $V$ under which $\mathcal{C}$ is the
    unique ground state for $|\mathcal{C}|$ points?

\item How simple can $V$ be?  Can it be decreasing?  Convex?

\item How large can the basin of attraction be made?
\end{enumerate}

In this paper, we give a complete answer to the first question, giving
necessary and sufficient conditions for such a potential to exist. The
second question is more subtle, but we rigorously answer it for several
important cases. In particular, we show that one can often use
remarkably simple potential functions.  The third question is the most
subtle of all, and there is little hope of providing rigorous proofs;
instead, experimental evidence must suffice.

The second question is particularly relevant for experimental work, for
example with colloids, because only a limited range of potentials can
be manipulated in the lab.  Inverse statistical mechanics with simple
potential functions was therefore raised as a challenge for future work
in ref.~\cite{To}.

\begin{figure}
\begin{center}
\includegraphics[scale=1.111]{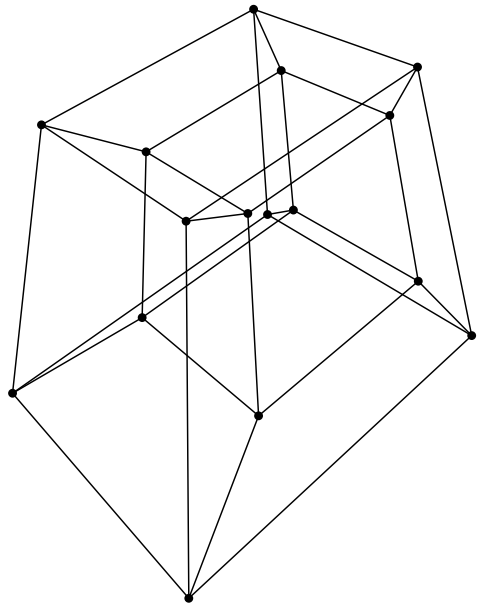}
\end{center}
\caption{The hypercube, drawn in four-point perspective.}
\label{fig:hypercube}
\end{figure}

\begin{figure}
\begin{center}
\includegraphics[scale=0.555]{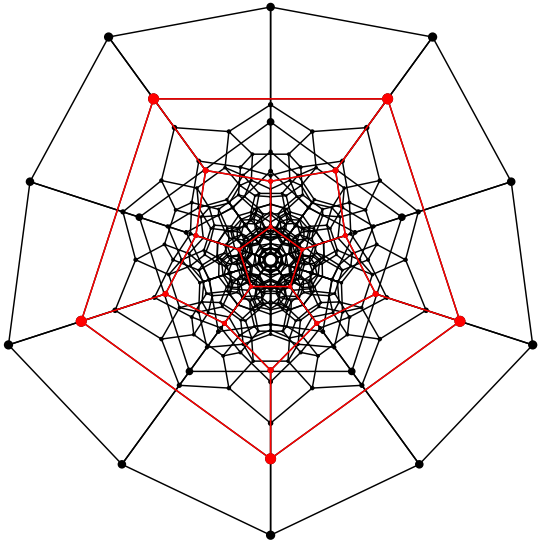}
\end{center}
\caption{The Schlegel diagram for the regular $120$-cell (with a
dodecahedral facet in red).}
\label{fig:120cell}
\end{figure}

We will focus on four especially noteworthy structures:
\begin{enumerate}
\item The $8$ vertices of a cube, with $6$ square facets.

\item The $20$ vertices of a regular dodecahedron, with $12$
    pentagonal facets.

\item The $16$ vertices of a hypercube in four dimensions, with $8$
    cubic facets (see Fig.~\ref{fig:hypercube}).

\item The $600$ vertices of a regular $120$-cell in four
    dimensions, with $120$ dodecahedral facets (see
    Fig.~\ref{fig:120cell}).
\end{enumerate}
The latter two configurations exist in four dimensions, but as
discussed above the sphere containing them is a three-dimensional
space, so they are intrinsically three-dimensional.

These configurations are important test cases, because they are elegant
and symmetrical yet at the same time not at all easy to build.  The
problem is that their facets are too large, which makes them highly
unstable.  Under ordinary potential functions, such as inverse power
laws, these configurations are never even local minima, let alone
global minima.  In the case of the cube, one can typically improve it
by rotating two opposite facets so they are no longer aligned.  That
lowers the energy, and indeed the global minimum appears to be the
antiprism arrived at via a $45^\circ$ rotation (and subsequent
adjustment of the edge lengths). It might appear that this process
always works, and that the cube can never minimize a convex, decreasing
potential function. However, careful calculation shows that this
argument is mistaken, and we will exhibit an explicit convex,
decreasing potential function for which the cube is provably the unique
global minimum.

One reason why the four configurations mentioned above are interesting
is that they are spherical analogues of the honeycomb and diamond
packings from $\R^2$ and $\R^3$, respectively.  In each of our four
cases, the nearest neighbors of any point form the vertices of a
regular spherical simplex.  They have the smallest possible
coordination numbers that can occur in locally jammed packings (see,
for example, ref.~\cite{DTSC}).

\section{Potential wells}

Suppose $\mathcal{C}$ is a configuration in $S^{n-1}$. The obvious way
to build $\mathcal{C}$ is to use deep potential wells (i.e., local
minima in $V$) corresponding to the distances between points in
$\mathcal{C}$, so that configurations that use only those distances are
energetically favored.  This method produces complicated potential
functions, which may be difficult to produce in the real world, but it
is systematic and straightforward. In this section we rigorously
analyze the limitations of this method and determine exactly when it
works, thereby answering a question raised towards the end of
ref.~\cite{RST1}.

The first limitation is obvious.  Define the \emph{distance
distribution} of $\mathcal{C}$ to be the function $d$ such that $d(r)$
is the number of pairs of points in $\mathcal{C}$ at distance $r$.  The
distance distribution determines the potential energy via
\begin{equation} \label{eq:EVlinear}
E_V(\mathcal{C}) = \sum_{r} V(r) \,d(r).
\end{equation}
Thus, $\mathcal{C}$ cannot possibly be the unique ground state unless
it is the only configuration with its distance distribution.

The second limitation is more subtle.  The formula~\eqref{eq:EVlinear}
shows that $E_V$ depends linearly on $d$.  If $d$ is a weighted average
of the distance distributions of some other configurations, then the
energy of $\mathcal{C}$ will be the same weighted average of the other
configurations' energies.  In that case, one of those configurations
must have energy at least as low as that of $\mathcal{C}$.  Call $d$
\emph{extremal} if it is an extreme point of the convex hull of the
space of all distance distributions of $|\mathcal{C}|$-point
configurations in $S^{n-1}$ (i.e., it cannot be written as a weighted
average of other distance distributions). If $\mathcal{C}$ is the
unique ground state for some isotropic pair potential, then $d$ must be
extremal.

For an example, consider three-point configurations on the circle
$S^1$, specified by the angles between the points (the shorter way
around the circle). The distance distribution of the configuration with
angles $45^\circ$, $90^\circ$, and $135^\circ$ is not extremal, because
it is the average of those for the $45^\circ$, $45^\circ$, $90^\circ$
and $90^\circ$, $135^\circ$, $135^\circ$ configurations.

\begin{theorem} \label{theorem:potwells}
If $\mathcal{C}$ is the unique configuration in $S^{n-1}$ with its
distance distribution $d$ and $d$ is extremal, then there exists a
smooth potential function $V \co (0,2] \to \R$ under which
$\mathcal{C}$ is the unique ground state among configurations of
$|\mathcal{C}|$ points in $S^{n-1}$.
\end{theorem}

The analogue of Theorem~\ref{theorem:potwells} for finite clusters of
particles in Euclidean space is also true, with almost exactly the same
proof.

{\it Proof:} Because $d$ is extremal, there exists a function $\ell$
defined on the support $\supp(d)$ of $d$ [i.e., the set of all $r$ such
that $d(r)\ne0$] such that $d$ is the unique minimum of
$$
t \mapsto \sum_r \ell(r)\, t(r)
$$
among $|\mathcal{C}|$-point distance distributions $t$ with $\supp(t)
\subseteq \supp(d)$.  Such a function corresponds to a supporting
hyperplane for the convex hull of the distance distributions with
support contained in $\supp(d)$.

For each $\varepsilon>0$, choose any smooth potential function
$V_\varepsilon$ such that $V_\varepsilon(r) = \ell(r)$ for $r \in
\supp(d)$ while
$$
V_\varepsilon(s) > \sum_{r} \ell(r)\, d(r)
$$
whenever $s$ is not within $\varepsilon$ of a point in $\supp(d)$. This
is easily achieved using deep potential wells, and it guarantees that
no configuration can minimize energy unless every distance occurring in
it is within $\varepsilon$ of a distance occurring in $\mathcal{C}$.
Furthermore, when $\varepsilon$ is sufficiently small [specifically,
less than half the distance between the closest two points in
$\supp(d)$], choose $V_\varepsilon$ so that for each $r \in \supp(d)$,
we have $V(s)
> V(r)$ whenever $|s-r| \le \varepsilon$ and $s \ne r$.

For a given $\varepsilon$, there is no immediate guarantee that
$\mathcal{C}$ will be the ground state.  However, consider what happens
to the ground states under $V_\varepsilon$ as $\varepsilon$ tends to
$0$. All subsequential limits of their distance distributions must be
distance distributions with support in $\supp(d)$.  Because of the
choice of $\ell$, the only possibility is that they are all $d$.  In
other words, as $\varepsilon$ tends to $0$ the distance distributions
of all ground states must approach $d$.  Because the number of points
at each given distance is an integer, it follows that when
$\varepsilon$ is sufficiently small, for each $r \in \supp(d)$, there
are exactly $d(r)$ distances in each ground state that are within
$\varepsilon$ of $r$.  Because $V_\varepsilon$ has a strict local
minimum at each point in $\supp(d)$, it follows that it is minimized at
$d$ (and only at $d$) when $\varepsilon$ is sufficiently small.  The
conclusion of the theorem then follows from our assumption that
$\mathcal{C}$ is the unique configuration in $S^{n-1}$ with distance
distribution $d$.  \qedbox

It would be interesting to have a version of
Theorem~\ref{theorem:potwells} for infinite collections of particles in
Euclidean space, but there are technical obstacles. Having infinitely
many distances between particles makes the analysis more complicated,
and one particular difficulty is what happens if the set of distances
has an accumulation point or is even dense (for example in the case of
a disordered packing).  In such a case there seems to be no simple way
to use potential wells, but in fact a continuous function with a
fractal structure can have a dense set of strict local minima, and
perhaps it could in theory serve as a potential function.

\section{Simulation-guided optimization}

In this section we describe an algorithm for optimizing the potential
function to create a specified ground state.  Our algorithm is similar
to, and inspired by, the zero-temperature optimization procedure
introduced in ref.~\cite{RST2}; the key difference is that their
algorithm is based on a fixed list of competing configurations and uses
simulated annealing, whereas ours dynamically updates that list and
uses linear programming.  (We also omit certain conditions on the
phonon spectrum that ensure mechanical stability.  In our algorithm,
they appear to be implied automatically once the list of competitors is
sufficiently large.)

Suppose the allowed potential functions are the linear combinations of
a finite set $V_1,\dots,V_k$ of specified functions.  In practice, this
may model a situation in which only certain potential functions are
physically realizable, with relative strengths that can be adjusted
within a specified range, while in theory we may choose the basic
potential functions so that their linear combinations can approximate
any reasonable function arbitrarily closely as $k$ becomes large.

Given a configuration $\mathcal{C} \subset S^{n-1}$, we wish to
choose a linear combination
$$
V = \lambda_1 V_1 + \dots + \lambda_k V_k
$$
so that $\mathcal{C}$ is the global minimum for $E_V$.  We may also
wish to impose other conditions on $V$, such as monotonicity or
convexity.  We assume that all additional conditions are given by
finitely many linear inequalities in the coefficients
$\lambda_1,\dots,\lambda_k$.  (For conditions such as monotonicity or
convexity, which apply over the entire interval $(0,2]$ of distances,
we approximate them by imposing these conditions on a large but finite
subset of the interval.)

Given a finite set of competitors
$\mathcal{C}_1,\dots,\mathcal{C}_\ell$ to $\mathcal{C}$, we can choose
the coefficients by solving a linear program.  Specifically, we add an
additional variable $\Delta$ and impose the constraints
$$
E_V(\mathcal{C}_i) \ge E_V(\mathcal{C}) +\Delta
$$
for $1 \le i \le\ell$, in addition to any additional constraints (as in
the previous paragraph).  We then choose $\lambda_1,\dots,\lambda_k$
and $\Delta$ so as to maximize $\Delta$ subject to these constraints.
Because this maximization problem is a linear program, its solution is
easily found.

If the coefficients can be chosen so that $\mathcal{C}$ is the global
minimum, then $\Delta$ will be positive and this procedure will produce
a potential function for which $\mathcal{C}$ has energy less than each
of $\mathcal{C}_1,\dots,\mathcal{C}_\ell$.  The difficulty is how to
choose these competitors.  In some cases, it is easy to guess the best
choices: for example, the natural competitors to a cube are the square
antiprisms.  In others, it is far from easy.  Which configurations
compete with the regular $120$-cell in $S^3$?

Our simulation-guided algorithm iteratively builds a list of
competitors and an improved potential function.  We start with any
choice of coefficients, say $\lambda_1=1$ and
$\lambda_2=\dots=\lambda_k=0$, and the empty list of competitors. We
then choose $|\mathcal{C}|$ random points on $S^{n-1}$ and minimize
energy by gradient descent to produce a competitor to $\mathcal{C}$,
which we add to the list (if it is different from $\mathcal{C}$) and
use to update the choice of coefficients. This alternation between
gradient descent and linear programming continues until either we are
satisfied that $\mathcal{C}$ is the global minimum of the potential
function, or we find a list of competitors for which linear programming
shows that $\Delta$ must be negative (in which case no choice of
coefficients makes $\mathcal{C}$ the ground state).

This procedure is only a heuristic algorithm.  When $\Delta$ is
negative, it proves that $\mathcal{C}$ cannot be the ground state
(using linear combinations of $V_1,\dots,V_k$ satisfying the desired
constraints), but otherwise nothing is proved.  As the number of
iterations grows large, the algorithm is almost certain to make
$\mathcal{C}$ the ground state if that is possible, because
eventually all possible competitors will be located.  However, we
have no bounds on the rate at which this occurs.

We hope that $\mathcal{C}$ will not only be the ground state, but will
also have a large basin of attraction under gradient descent.
Maximizing the energy difference $\Delta$ seems to be a reasonable
approach, but other criteria may do even better.  In practice,
simulation-guided optimization does not always produce a large basin of
attraction, even when one is theoretically possible.  Sometimes it
helps to remove the first handful of competitors from the list once the
algorithm has progressed far enough.

\section{Rigorous analysis}

The numerical method described in the previous section appears to work
well, but it is not supported by rigorous proofs. In this section we
provide such proofs in several important cases.  The key observation is
that the conditions for proving a sharp bound in
Proposition~\ref{prop:yudin} below are themselves linear and can be
added as constraints in the simulation-guided optimization.  While this
does not always lead to a solution, when it does the solution is
provably optimal (and in fact no simulations are then needed).  The
prototypical example is the following theorem:

\begin{theorem}
\label{theorem:cube} Let the potential function $V \co (0,2] \to \R$
be defined by
$$
V(r) = \frac{1}{r^3} - \frac{1.13}{r^6} + \frac{0.523}{r^9}.
$$
Then the cube is the unique global energy minimum among $8$-point
configurations on $S^2$.  The function $V$ is decreasing and
strictly convex.
\end{theorem}

Theorem~\ref{theorem:cube} is stated in terms of a specific potential
function, but of course many others could be found using our algorithm.
Furthermore, as discussed in the conclusions below, the proof
techniques are robust and any potential function sufficiently close to
this one works.

The potential function used in Theorem~\ref{theorem:cube} is modeled
after the Lennard-Jones potential.  The simplest generalization (namely
a linear combination of two inverse power laws) cannot work here, but
three inverse power laws suffice.  The potential function in
Theorem~\ref{theorem:cube} is in fact decreasing and convex on the
entire right half-line, although only the values on $(0,2]$ are
relevant to the problem at hand and the potential function could be
extended in an arbitrary manner beyond that interval.

To prove Theorem~\ref{theorem:cube}, we will apply \emph{linear
programming bounds}, in particular, Yudin's version for potential
energy \cite{Y}.  Let $P_i$ denote the $i$th degree Gegenbauer
polynomial for $S^{n-1}$ [i.e., with parameter $(n-3)/2$, which we
suppress in our notation for simplicity], normalized to have
$P_i(1)=1$.  These are a family of orthogonal polynomials that arise
naturally in the study of harmonic analysis on $S^{n-1}$. The
fundamental property they have is that for every finite configuration
$\mathcal{C} \subset S^{n-1}$,
$$
\sum_{x,y\in \mathcal{C}} P_i(\langle x,y\rangle) \ge 0.
$$
(Here, $\langle x,y \rangle$ denotes the inner product, or dot product,
between $x$ and $y$.) See section~2.2 of ref.~\cite{CK} for further
background.

The linear programming bound makes use of an auxiliary function $h$ to
produce a lower bound on potential energy.  The function $h$ will be a
polynomial $ h(t) = \sum_{i=0}^d \alpha_i P_i(t) $ with coefficients
$\alpha_0,\dots,\alpha_d \ge 0$.  It will also be required to satisfy $
h(t) \le V\big(\sqrt{2-2t}\big) $ for all $t \in [-1,1)$.  Note that
$\sqrt{2-2t}$ is the Euclidean distance between two unit vectors with
inner product $t$, because $|x-y|^2 = |x|^2+|y|^2-2\langle x,y \rangle
= 2 - 2\langle x,y \rangle$ when $|x|=|y|=1$. We view $h$ as a function
of the inner product, and the previous inequality simply says that it
is a lower bound for $V$.

We say the configuration $\mathcal{C}$ is \emph{compatible} with $h$ if
two conditions hold.  The first is that $h(t) = V\big(\sqrt{2-2t}\big)$
whenever $t$ is the inner product between two distinct points in
$\mathcal{C}$.  The second is that whenever $\alpha_i>0$ with $i>0$, we
have $ \sum_{x,y\in \mathcal{C}} P_i(\langle x,y\rangle) = 0. $ This
equation holds if and only if for every $z \in S^{n-1}$, $ \sum_{x\in
\mathcal{C}} P_i(\langle x,z\rangle) = 0. $ (The subtle direction
follows from Theorem~9.6.3 in ref.~\cite{AAR}.)

\begin{proposition}[Yudin \cite{Y}] \label{prop:yudin}
Given the hypotheses listed above for $h$, every $N$-point
configuration in $S^{n-1}$ has $V$-potential energy at least
$(N^2\alpha_0 - Nh(1))/2. $ If $\mathcal{C}$ is compatible with $h$,
then it is a global minimum for energy among all $|\mathcal{C}|$-point
configurations in $S^{n-1}$, and every such global minimum must be
compatible with $h$.
\end{proposition}

{\it Proof:} Let $\mathcal{C} \subset S^{n-1}$ be any finite
configuration with $N$ points (not necessarily compatible with $h$).
Then
\begin{align*}
E_V(\mathcal{C})
&= \frac{1}{2}\sum_{x,y\in \mathcal{C},\, x \ne y} V
\big(\sqrt{2-2\langle x,y \rangle}\big)\\
& \ge  \frac{1}{2}\sum_{x,y\in \mathcal{C},\, x \ne y} h
\big(\langle x,y \rangle\big)\\
& =  -\frac{Nh(1)}{2} + \frac{1}{2}\sum_{x,y\in \mathcal{C}} h
\big(\langle x,y \rangle\big)\\
& =  -\frac{Nh(1)}{2} + \frac{1}{2}\sum_{i=0}^d \alpha_i
\sum_{x,y\in \mathcal{C}} P_i
\big(\langle x,y \rangle\big)\\
& \ge  -\frac{Nh(1)}{2} + \frac{\alpha_0}{2} \sum_{x,y\in
\mathcal{C}} P_0 \big(\langle x,y \rangle\big)\\
& = \frac{N^2\alpha_0 - Nh(1)}{2}.
\end{align*}
The first inequality holds because $h$ is a lower bound for $V$, and
the second holds because all the $P_i$-sums are nonnegative (as are the
coefficients $\alpha_i$).  The lower bound for energy is attained by
$\mathcal{C}$ if and only if both inequalities are tight, which holds
if and only if $\mathcal{C}$ is compatible with $h$, as desired.
\qedbox

{\it Proof of Theorem~\ref{theorem:cube}:} It is straightforward to
check that $V$ is decreasing and strictly convex. To prove that the
cube is the unique local minimum, we will use linear programming
bounds.

Let $h$ be the unique polynomial of the form
$$
h(t) = \alpha_0 + \alpha_1 P_1(t) + \alpha_2 P_2(t) + \alpha_3
P_3(t) + \alpha_5 P_5(t)
$$
(note that $P_4$ is missing) such that $h(t)$ agrees with
$V\big(\sqrt{2-t}\big)$ to order $2$ at $t=\pm 1/3$ and to order $1$ at
$t=-1$.  These values of $t$ are the inner products between distinct
points in the cube.  One can easily compute the coefficients of $h$ by
solving linear equations and verify that they are all positive.
Furthermore, it is straightforward to check that $h(t) \le
V\big(\sqrt{2-2t}\big)$ for all $t \in [-1,1)$, with equality only for
$t \in \{-1,-1/3,1/3\}$.

The cube is compatible with $h$, and to complete the proof all that
remains is to show that it is the only $8$-point configuration that
is compatible with $h$.  Every such configuration $\mathcal{C}$ can
have only $-1$, $-1/3$, and $1/3$ as inner products between distinct
points.  For each $y \in \mathcal{C}$ and $1 \le i \le 3$,
$$
\sum_{x \in \mathcal{C}} P_i(\langle x,y \rangle) = 0.
$$
If there are $N_t$ points in $\mathcal{C}$ that have inner product
$t$ with $y$, then
$$
P_i(-1) N_{-1} + P_i(-1/3) N_{-1/3} + P_i(1/3) N_{1/3} + P_i(1) = 0
$$
for $1 \le i \le 3$.  These linear equations have the unique
solution $N_{-1}=1$, $N_{\pm1/3} = 3$.

In other words, not only is the complete distance distribution of
$\mathcal{C}$ determined, but the distances from each point to the
others are independent of which point is chosen.  The remainder of the
proof is straightforward.  For each point in $\mathcal{C}$, consider
its three nearest neighbors.  They must have inner product $-1/3$ with
each other: no two can be antipodal to each other, and if any two were
closer together than in a cube, then some other pair would be farther
(which is impossible).  Thus, the local configuration of neighbors is
completely determined, and in this case, that determines the entire
structure. \qedbox

\begin{theorem}
\label{theorem:dodecahedron} Let the potential function $V \co (0,2]
\to \R$ be defined by
$$
V(r) =
(1+t)^5+\frac{(t+1)^2(t-1/3)^2(t+1/3)^2(t^2-5/9)^2}{6(1-t)^2},
$$
where $t = 1-r^2/2$. Then the regular dodecahedron is the unique global
energy minimum among $20$-point configurations on $S^2$.  The function
$V$ is decreasing and strictly convex.
\end{theorem}

The proof is analogous to that of Theorem~\ref{theorem:cube}, except
that we choose $h(t) = (1+t)^5.$ The proof of uniqueness works
similarly.  Note that the potential function used in
Theorem~\ref{theorem:dodecahedron} is physically unnatural.  It does
not seem worth carefully optimizing the form of this potential function
when it is already several steps away from real-world application.
Instead, Theorems~\ref{theorem:dodecahedron}
through~\ref{theorem:120cell} should be viewed as plausibility
arguments, which prove that there exists a convex, decreasing potential
while allowing its form to be highly complicated.

\begin{theorem}
\label{theorem:hypercube} Let the potential function $V \co (0,2] \to
\R$ be defined by
\begin{align*}
V(r) = {}&
\frac{-13+73t+5t^2+7t^3+t^5}{120}\\
&+\frac{7(t+1)(t+1/2)^2t^2(t-1/2)^2}{120(1-t)},
\end{align*}
where $t = 1-r^2/2$. Then the hypercube is the unique global energy
minimum among $16$-point configurations on $S^3$.  The function $V$ is
decreasing and strictly convex.
\end{theorem}

For the $120$-cell, let $q(t)$ be the monic polynomial whose roots
are the inner product between distinct points in the $120$-cell;
in other words,
\begin{align*}
q(t) = {}& t (t+1) (4t+1)(4t-1) (4t+3)(4t-3)\\
& \cdot (2t+1)(2t-1)(16t^2-5)\\
& \cdot (4t^2+2t-1) (4t ^2-2t-1)\\
& \cdot (16t^2+4t-1) (16t^2-4t-1)\\
& \cdot (16t^2+4t-11)(16 t^2-4t-11)\\
& \cdot (16t^2+12t+1)(16t^2-12t+1)\\
& \cdot (16t^2+20t+5)(16t^2-20t+5)/2^{50}.
\end{align*}

Let $m_1,\dots,m_{29}$ be the integers $2$, $4$, $6$, $8$, $10$, $14$,
$16$, $18$, $22$, $26$, $28$, $34$, $38$, $46$, $1$, $3$, $5$, $7$,
$9$, $11$, $13$, $15$, $17$, $19$, $21$, $23$, $25$, $27$, $29$ (in
order), and let $c_1,\dots,c_{17}$ be $1$, $2/3$, $4/9$, $1/4$, $1/9$,
$1/20$, $1/20$, $1/15$, $1/15$, $9/200$, $3/190$, $0$, $7/900$, $1/40$,
$1/35$, $3/190$, and $1/285$.

\begin{theorem}
\label{theorem:120cell} Let the potential function $V \co (0,2] \to
\R$ be defined by
$$
V(r) = \sum_{i=1}^{17} c_i P_i(t) + \sum_{i=1}^{29}
\frac{P_{m_i}(t)}{10^{6}} + 10^5\frac{q(t)^2}{1-t},
$$
where $t=1-r^2/2$. Then the regular $120$-cell is the unique global
energy minimum among $600$-point configurations on $S^3$.  The function
$V$ is decreasing and strictly convex.
\end{theorem}

The proofs of Theorems~\ref{theorem:hypercube}
and~\ref{theorem:120cell} use the same techniques as before.  The most
elaborate case is the $120$-cell, specifically the proof of uniqueness.
The calculation of the coefficients $N_t$, as in the proof of
Theorem~\ref{theorem:cube}, proceeds as before, except that the
$P_{12}$ sum does not vanish (note that the coefficient $c_{12}$ of
$P_{12}$ in $V$ is zero).  Nevertheless, there are enough simultaneous
equations to calculate the numbers $N_t$.  Straightforward case
analysis suffices to show then that the four neighbors of each point
form a regular tetrahedron, and the entire structure is determined by
that.

\section{Conclusions and open problems}

In this article, we have shown that symmetrical configurations can
often be built by using surprisingly simple potential functions, and we
have given a new algorithm to search for such potential functions.
However, many open problems remain.

One natural problem is to extend the linear programming bound analysis
to Euclidean space.  There is no conceptual barrier to this (section~9
of ref.~\cite{CK} develops the necessary theory), but there are
technical difficulties that must be overcome if one is to give a
rigorous proof that a ground state has been achieved.

A second problem is to develop methods of analyzing the basin of
attraction of a given configuration under gradient descent.  We know of
no rigorous bounds for the size of the basin.

The review article \cite{To} raises the issue of robustness: Will a
small perturbation in the potential function (due, for example, to
experimental error) change the ground state?  One can show that the
potential in Theorem~\ref{theorem:cube} is at least somewhat robust.
Specifically, it follows from the same proof techniques that there
exists an $\varepsilon>0$ such that if the values of the potential
function and its first two derivatives are changed by a factor of no
more than $1+\varepsilon$, then the ground state remains the same. It
would be interesting to see how robust a potential function one could
construct in this case.  The argument breaks down slightly for
Theorems~\ref{theorem:dodecahedron} through~\ref{theorem:120cell}, but
they can be slightly modified to make them robust.

\begin{figure}
\begin{center}
\includegraphics{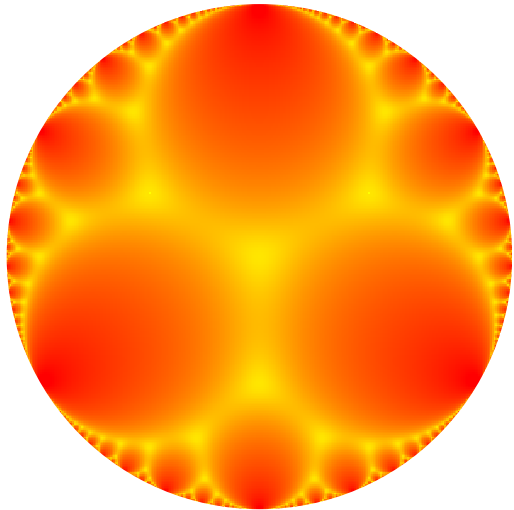}
\end{center}
\caption{Gaussian energy on the space of two-dimensional lattices
(red means high energy).}
\label{fig:poincare}
\end{figure}

It would also be interesting to develop a clearer geometrical picture
of energy minimization problems.  For example, for Bravais lattices in
the plane, the space of lattices can be naturally described by using
hyperbolic geometry [see, for example, ref.~\cite{T}, (pp~124--125)].
Fig.~\ref{fig:poincare} shows a plot of potential energy for a Gaussian
potential function, drawn by using the Poincar\'e disk model of the
hyperbolic plane.  Each point corresponds to a lattice, and the color
indicates energy (red is high).  The local minima in yellow are copies
of the triangular lattice; the different points correspond to different
bases. The saddle points between them are square lattices, which can
deform into triangular lattices in two different ways by shearing the
square along either axis.  The red points on the boundary show how the
energy blows up as the lattice becomes degenerate.  In more general
energy minimization problems, we cannot expect to draw such pictures,
but one could hope for a similarly complete analysis, with an
exhaustive list of all critical points as well as a description of how
they are related to each other geometrically.

\section*{Acknowledgements}

We are grateful to Salvatore Torquato for helpful discussions as well
as comments on our manuscript and the anonymous referees for useful
feedback. Kumar was partially supported by National Science Foundation
grant DMS-0757765.


\begin{thebibliography}{99}

\bibitem{To} Torquato S (2009) Inverse optimization techniques
    for targeted self-assembly.  \textit{Soft Matter\/} 5:1157--1173,
    \texttt{arXiv:0811.0040}.

\bibitem{RST1} Rechtsman M, Stillinger FH, Torquato S (2005)
    Optimized interactions for targeted self-assembly:
    application to honeycomb lattice.  \textit{Phys Rev Lett\/}
    95:228301, \texttt{arXiv:cond-mat/0508495}.

\bibitem{RST2} Rechtsman M, Stillinger FH, Torquato S (2006)
    Designed isotropic potentials via inverse methods for
    self-assembly. \textit{Phys Rev E\/} 73:011406,
    \texttt{arXiv:cond-mat/0603415}.

\bibitem{RST3} Rechtsman M, Stillinger FH, Torquato S (2006)
    Self-assembly of the simple cubic lattice via an isotropic
    potential. \textit{Phys Rev E\/} 74:021404,
    \texttt{arXiv:cond-mat/0606674}.

\bibitem{RST4} Rechtsman M, Stillinger FH, Torquato S (2007)
    Synthetic diamond and Wurtzite structures self-assemble
    with isotropic pair interactions. \textit{Phys Rev E\/}
    75:031403, \texttt{arXiv:0709.3807}.

\bibitem{L} Likos CN (2001) Effective interactions in soft
    condensed matter physics. \textit{Phys Rep\/} 348:267--439.

\bibitem{RSS} Russel WB, Saville DA, Showalter WR (1989)
    \textit{Colloidal Dispersions\/} (Cambridge Univ Press,
    Cambridge).

\bibitem{SM} Sadoc J-F, Mosseri R (1999) \textit{Geometrical
    Frustration\/} (Cambridge Univ Press, Cambridge).

\bibitem{DHNMBW} Dinsmore AD, et al.\ (2002) Colloidosomes:
    selectively permeable capsules
    composed of colloidal particles. \textit{Science\/} 298:1006--1009.

\bibitem{BG}  Bowick M, Giomi L (2009) Two-dimensional
    matter: order, curvature and defects. \textit{Adv Phys\/},
    \texttt{arXiv:\break0812.3064v1 [cond-mat.soft]}.

\bibitem{DTSC} Donev A, Torquato S, Stillinger FH, Connelly R
    (2004) Jamming in hard sphere and disk packings. \textit{J Appl
    Phys\/} 95:989--999.

\bibitem{Y} Yudin VA (1993) Minimum potential energy of a
    point system of charges.
    \textit{Discrete Math Appl\/} 3:75--81.

\bibitem{CK}  Cohn H, Kumar A (2007) Universally optimal
    distribution of points on spheres. \textit{J Am Math Soc\/} 20:99--148,
    \texttt{arXiv:math.MG/0607446}.

\bibitem{AAR} Andrews G, Askey R, Roy R (1999) \textit{Special
    Functions\/} (Cambridge Univ Press, Cambridge).

\bibitem{T} Terras A (1985) \textit{Harmonic Analysis on Symmetric
    Spaces
    and Applications\/} (Springer, New York), Vol.~I.

\end{thebibliography}
\end{document}